\documentclass[twocolumn]{jpsj3}
\usepackage{txfonts}
\usepackage{graphicx}
\usepackage{epstopdf}
\usepackage{xcolor}

\title{Evolution of Superconductivity in Fe$_{1+y}$Te$_{1-x}$Se$_{x}$ Annealed in Te Vapor }

\author{Yue Sun$^{1,2}$, Yuji Tsuchiya$^2$, Tatsuhiro Yamada$^2$, Toshihiro Taen$^2$, Sunseng Pyon$^2$, Zhixiang Shi$^{1*}$, and Tsuyoshi Tamegai$^{2\dag}$}
\inst{$^1$Department of Physics, Southeast University, Nanjing 211189, People's Republic of China \\
$^2$Department of Applied Physics, The University of Tokyo, 7-3-1 Hongo, Bunkyo-ku, Tokyo 113-8656, Japan} 

\abst{We report a detailed study of the Te vapor annealing effect in Fe$_{1+y}$Te$_{0.6}$Se$_{0.4}$ single crystals. Bulk superconductivity can be gradually induced by annealing with increasing amount of Te, until the molar ratio of Te to the sample reaches 1 : 10. By further increasing Te molar ratio, superconducting volume is gradually reduced. Resistivity and Hall effect measurements manifest that annealing in Te vapor delocalizes the charge carriers by removing excess Fe. The optimally Te annealed crystal shows homogeneous critical current densities with a large value of $\sim$ 4.3 $\times$ 10$^5$ A/cm$^2$, which proves that the Te annealing is effective to induce bulk superconductivity in Fe$_{1+y}$Te$_{1-x}$Se$_{x}$.}

\kword{Fe$_{1+y}$Te$_{1-x}$Se$_{x}$, Te annealing effect, iron chalcogenide, excess Fe, Hall effect, critical current density}

\begin{document}
\maketitle
Following the discovery of superconductivity at 26 K in an iron oxypnictide LaFeAs(O, F),\cite{1} research activities of iron-based superconductors (IBSs) have been extended rapidly to a widen variety of materials.\cite{2} Among these, iron chalcogenides attracted much more attention recently due to the unexpected high \emph{T}$_c$. Although the \emph{T}$_c$ in FeSe is only 8 K,\cite{3} it increases to 14 K\cite{4, 5} with appropriate Te substitution for Se and 37 K\cite{6, 7} under high pressure. Furthermore, by intercalating spacer layers between adjacent FeSe layers, \emph{T}$_c$ reaches $\sim$ 32 K\cite{8} in \emph{A}$_x$Fe$_{2-y}$Se$_2$ (\emph{A} = K, Cs, Rb and Tl) and 43 K\cite{9} in Li$_x$(NH$_2$)$_y$(NH$_3$)$_{1-y}$Fe$_2$Se$_2$ (\emph{x} $\sim$ 0.6; \emph{y} $\sim$ 0.2). By applying pressure, \emph{T}$_c$ for \emph{A}$_x$Fe$_{2-y}$Se$_2$ can even reach $\sim$ 48 K.\cite{10} Recent angle-resolved photoemission spectroscopy (ARPES) reveals an unexpected large superconducting gap $\sim$ 19 meV in a single layered FeSe, which suggests a \emph{T}$_c$ as high as 65 K.\cite{11} Among iron chalcogenides, Fe$_{1+y}$Te$_{1-x}$Se$_{x}$ is unique in their structural simplicity, composed of only iron-chalcogenide layers, which is preferable for probing the mechanism of superconductivity. And its less toxic nature is also advantage for application of IBSs. However, superconductivity and magnetism of this compound are not only dependent on the doping level, but also sensitively to Fe non-stoichiometry, which originates from the partial occupation of excess Fe at the interstitial site in the Te/Se layer.\cite{12, 13} For the undoped parent compound Fe$_{1+y}$Te, the commensurate antiferromagnetic order can be tuned by the excess Fe to an incommensurate magnetic structure.\cite{12} In Se-doped FeTe samples, excess Fe was found to suppress superconductivity and cause the magnetic correlations.\cite{13} In the other end member Fe$_{1+y}$Se, superconductivity is reported to reside only in a very narrow concentration region of excess Fe.\cite{14} The Fe non-stoichiometry leads to many controversies in the mechanism of superconductivity in this system and also hinders its application progress.\cite{15, 16}

To remove the effect of Fe nonstoichiometry, some attempts have been made by annealing crystals under different conditions. Our previous report systematically compared different annealing conditions like vacuum, N$_2$, O$_2$, I$_2$ atmosphere, and immersing samples into dilute acid and alcoholic beverages. Through these investigations, O$_2$ annealing was proved to be the most effective way to induce bulk superconductivity in Fe$_{1+y}$Te$_{1-x}$Se$_{x}$ single crystals.\cite{17} Recently Koshika et al.\cite{18} reported superconductivity in Fe$_{1+y}$Te$_{1-x}$Se$_{x}$ can also be induced by annealing in Te vapor. In this paper, we reported the evolution of superconductivity in Fe$_{1+y}$Te$_{0.6}$Se$_{0.4}$ single crystals under annealing in controlled amount of Te vapor. Our results confirm that superconductivity can be induced by annealing in Te vapor. Furthermore, the well annealed sample shows a large and homogeneous critical current density, which proves the Te vapor annealing is another effective way to induce bulk superconductivity in Fe$_{1+y}$Te$_{1-x}$Se$_{x}$.

Single crystal with a nominal composition FeTe$_{0.6}$Se$_{0.4}$ was grown by the self-flux method as reported before.\cite{17} The obtained crystals were then cut and cleaved into thin slices with dimensions about 2.0 $\times$ 1.0 $\times$ 0.05 mm$^3$, weighed and loaded into a quartz tube (\emph{d} $\sim$ 10 mm $\phi$) with appropriate amount of Te grains. The quartz tube was carefully evacuated by a diffusion pump, and sealed into a length of 100 mm. During these processes, a diaphragm-type manometer with accuracy better than 1 mTorr was used for real-time monitoring the pressure in the system to prevent gas leakage. Then the crystals were annealed at 400 $^{\circ}$C for 20 h, followed by water quenching. Magnetization measurements were performed using a commercial superconducting quantum interference device (SQUID). Structure of the surface layers of the annealed crystal was characterized by means of X-ray diffraction (XRD) with Cu-\emph{K}$\alpha$ radiation. Microstructural and compositional investigations of the sample were performed using a scanning electron microscope (SEM) equipped with an energy dispersive x-ray spectroscopy (EDX).  Longitudinal and transverse (Hall) resistivities were measured by six-lead method with a Physical Property Measurement System (PPMS, Quantum Design) at temperatures down to 2 K and magnetic fields up to 9 T. Magneto-optical (MO) images were obtained by using the local field-dependent Faraday effect in the in-plane magnetized garnet indicator film employing a differential method.\cite{19}
\begin{figure}\center

　　\includegraphics[width=8cm]{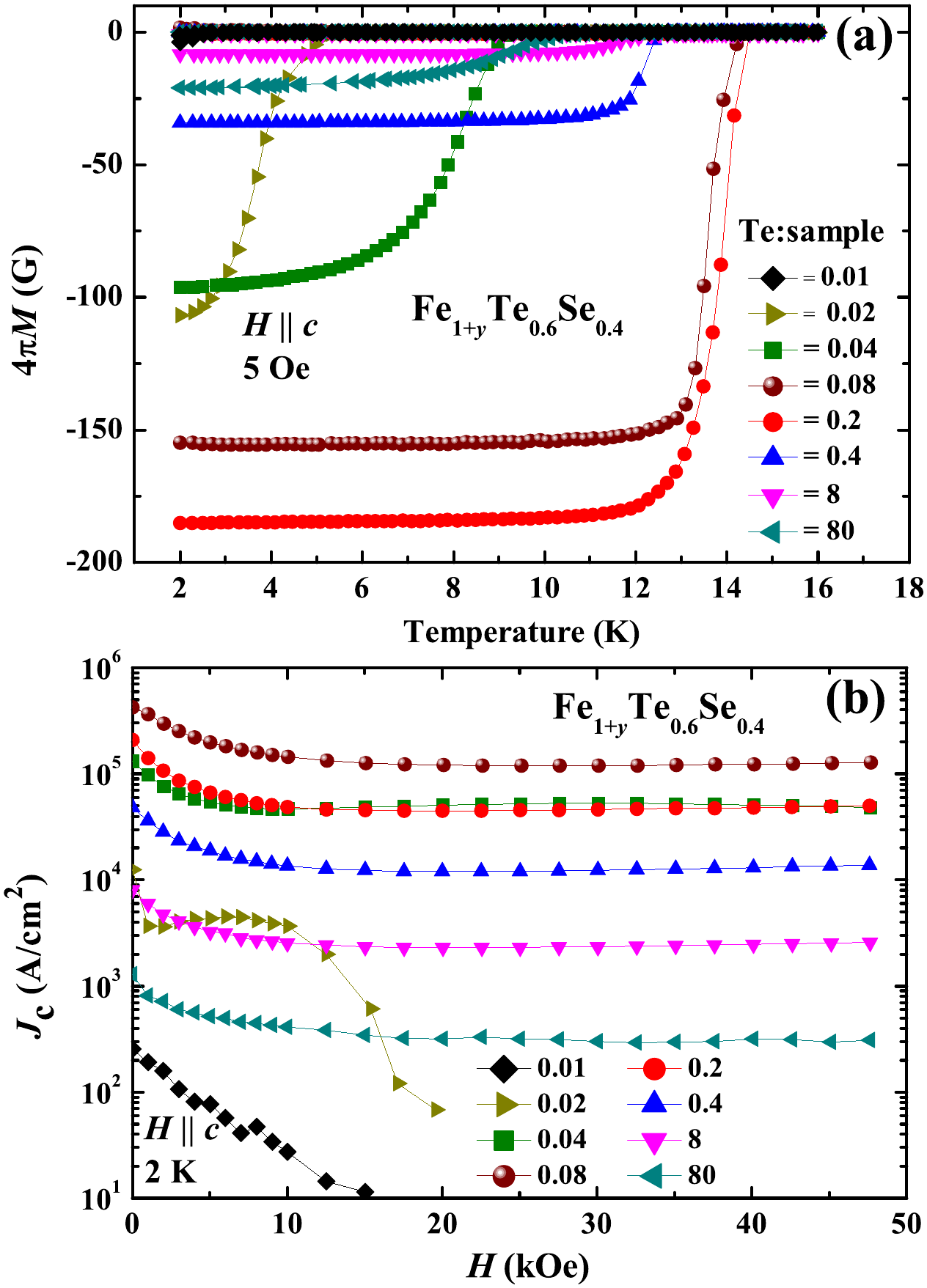}\\
　　\caption{(Color online) (a) Temperature dependence of zero-field-cooled (ZFC) and field-cooled (FC) magnetization at 5 Oe for Fe$_{1+y}$Te$_{0.6}$Se$_{0.4}$ single crystals annealed at 400 $^{\circ}$C with increasing amount of Te vapor (molar ratio of Te to the sample ranging from 0.01 to 80). (b) Magnetic field dependence of \emph{J}$_c$ at 2 K for Fe$_{1+y}$Te$_{1-x}$Se$_{x}$ annealing with increasing amount of Te vapor.}\label{}
\end{figure}

Fig. 1(a) shows the temperature dependence of zero-field-cooled (ZFC) and field-cooled (FC) magnetization at 5 Oe for Fe$_{1+y}$Te$_{0.6}$Se$_{0.4}$ single crystal annealed with increasing amount of Te (molar ratio of the Te to the sample ranging from 1 :100 to 80 : 1). As-grown crystal usually shows no superconductivity or very weak diamagnetic signal below 3 K. Annealing with increasing amount of Te, \emph{T}$_c$ is gradually enhanced, together with an increase in diamagnetic signal. After the ratio exceeds 0.1 - 0.2, \emph{T}$_c$ and the diamagnetic signal is suppressed with annealing. Actually, it is well known that even the sample is mostly non-superconducting, the diamagnetic signal can be significant when the non-superconducting region is covered by superconducting layer. Thus, to probe the evolution of superconductivity during the annealing, we refer to the critical current density \emph{J}$_c$, which is calculated from magnetic hysteresis loops (MHLs). From the MHLs, we can obtain \emph{J}$_c$ by using the Bean model\cite{20}:
$$ J_c=20\Delta M/(a(1-a/3b)) \eqno{(1)}$$
where $\Delta$\emph{M} is \emph{M}$_{down}$ - \emph{M}$_{up}$, \emph{M}$_{up}$ and \emph{M}$_{down}$ are the magnetizations when sweeping fields up and down, respectively, \emph{a} and \emph{b} are sample widths (\emph{a} $<$ \emph{b}). Magnetic field dependence of \emph{J}$_c$ at 2 K with increasing Te content is summarized in Fig. 1(b). It is obvious that samples annealed with less Te vapor (such as the ratio of 0.01 and 0.02) show smaller \emph{J}$_c$, and it is easily suppressed at high fields. The value of \emph{J}$_c$ is enhanced quickly with increasing amount of Te, reaching the maximum value at the ratio $\sim$ 0.1. Along with the increase in the value of \emph{J}$_c$, it also becomes almost independent of applied field. By further increasing the Te content, \emph{J}$_c$ is suppressed while keeping its field-independence.
\begin{figure}\center

　　\includegraphics[width=8cm]{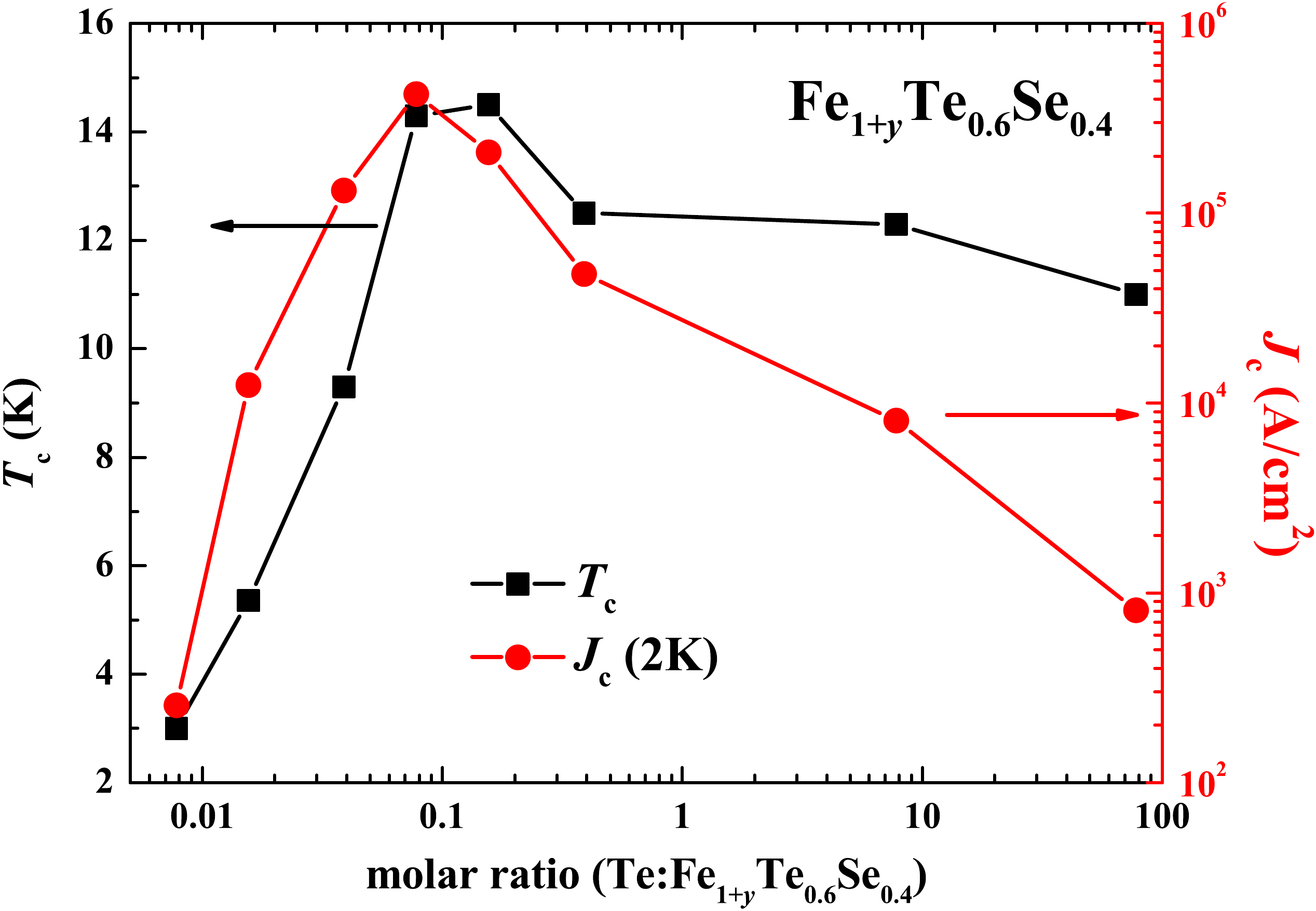}\\
　　\caption{(Color online) \emph{T}$_c$ and self-field \emph{J}$_c$(2 K) as functions of molar ratio of Te to Fe$_{1+y}$Te$_{0.6}$Se$_{0.4}$ single crystal annealed at 400 $^{\circ}$C.}\label{}
\end{figure}
To clearly show the evolution of superconductivity under annealing with Te vapor, we summarize \emph{T}$_c$ and \emph{J}$_c$(2 K) as a function of the amount of Te in Fig. 2. It is clear that both \emph{T}$_c$ and \emph{J}$_c$ are gradually enhanced by annealing with increasing amount of Te, and reach the optimal values of \emph{T}$_c$ $\sim$ 14.3 K and \emph{J}$_c$ $\sim$ 4.3 $\times$ 10$^5$ A/cm$^2$ when the molar ratio of Te to the sample is $\sim$ 0.1. After that, both \emph{T}$_c$ and \emph{J}$_c$ decrease with increasing the amount of Te. Now we focus on the \emph{J}$_c$(2 K), which manifests the evolution of bulk superconductivity as we pointed out before. The increasing and decreasing parts of \emph{J}$_c$ are asymmetric indicating different mechanisms. The increment of \emph{J}$_c$ comes from the increasing annealed volume of the sample. When the whole crystal is totally annealed to become superconducting, remaining Te will react with the crystal itself, which reduces the superconducting volume. It can also explain that although the \emph{J}$_c$ decreases after the optimal condition, it keeps magnetic field-independence.

\begin{figure}\center

　　\includegraphics[width=8cm]{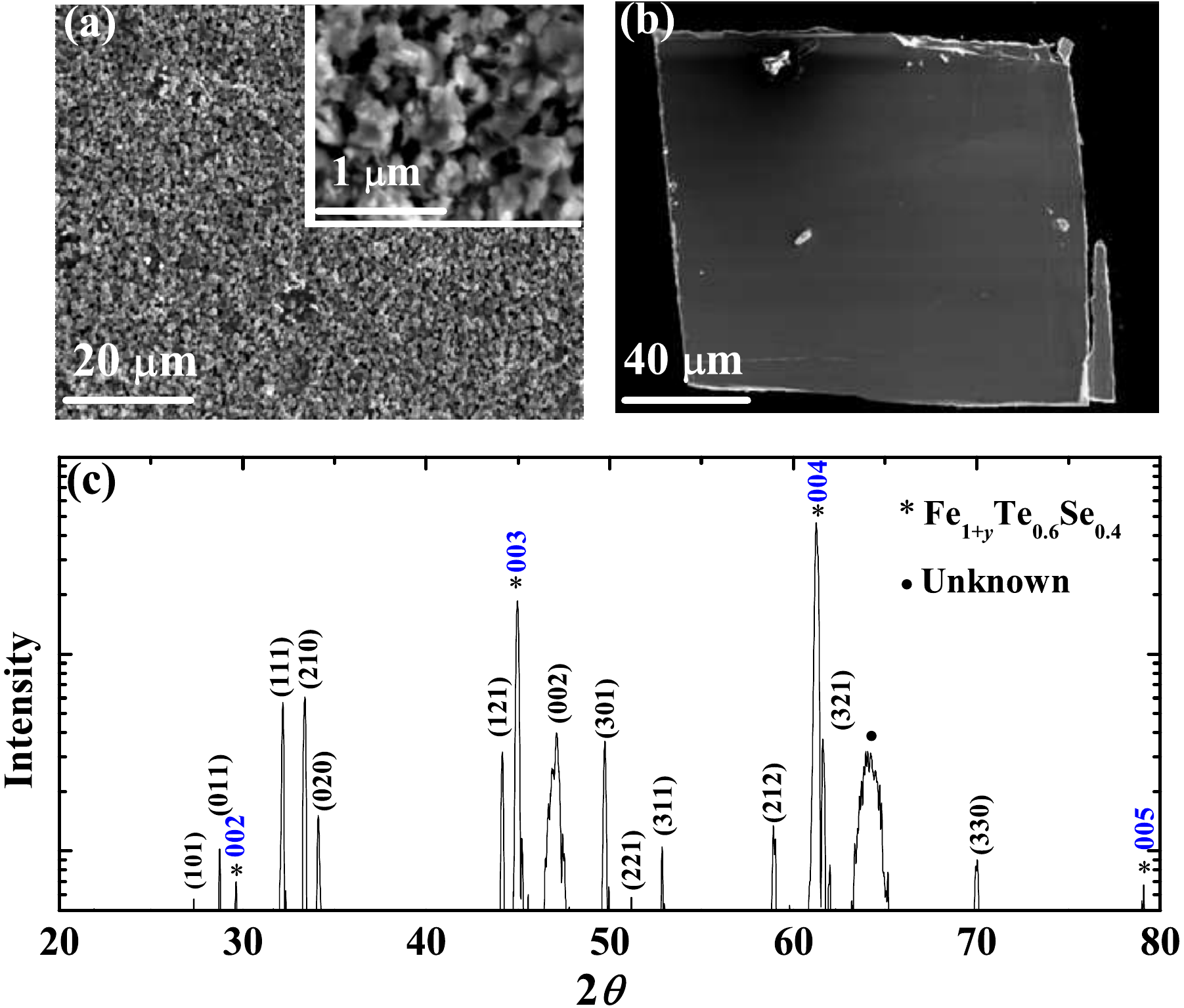}\\
　　\caption{(Color online) (a) Scanning electron microscope (SEM) images for annealed Fe$_{1+y}$Te$_{0.6}$Se$_{0.4}$ single crystal. Inset is an enlarged region. (b) SEM image for annealed crystal after cleaving the surface layers. (c) X-ray diffraction (XRD) pattern of the polycrystal-like surface for the annealed sample. Miller indices in brackets are for FeTe$_2$. }\label{}
\end{figure}
Fig. 3(a) shows the SEM images of the annealed crystal (molar ratio of Te to the sample is $\sim$ 0.08; the same below). After annealing in Te vapor, surface layers of the single crystal turn into polycrystal-like. EDX measurements show that the polycrystal-like surface contains only Fe and Te with a molar ratio roughly 1 : 2. XRD pattern of the polycrystal-like surface is shown in Fig. 3(c), in which almost all the main peaks can be identified by the structure of FeTe$_2$,\cite{21} except for (\emph{00l}) peaks from Fe$_{1+y}$Te$_{0.6}$Se$_{0.4}$ single crystal. The peak positions of FeTe$_2$ are slightly  shifted from the literature values possibly due to the Fe non-stoichiometry.\cite{22} A similar result has been reported by Koshika et al.\cite{18} Thus, the EDX and XRD results show the polycrystal-like surface is FeTe$_2$. The FeTe$_2$ surface layers in the annealed crystal may come from the reaction between excess Fe with the Te vapor, although we cannot simply rule out the possibility of Fe in the regular site also reacted with Te vapor. Fig. 3(b) shows the SEM image of the inner part of the annealed crystal after cleaving the polycrystal-like surface. The inner part of the crystal still keeps the mirror-like surface, and EDX shows that the inner surface is composed of Fe$_{1+y}$Te$_{0.6}$Se$_{0.4}$.

\begin{figure}\center

　　\includegraphics[width=8cm]{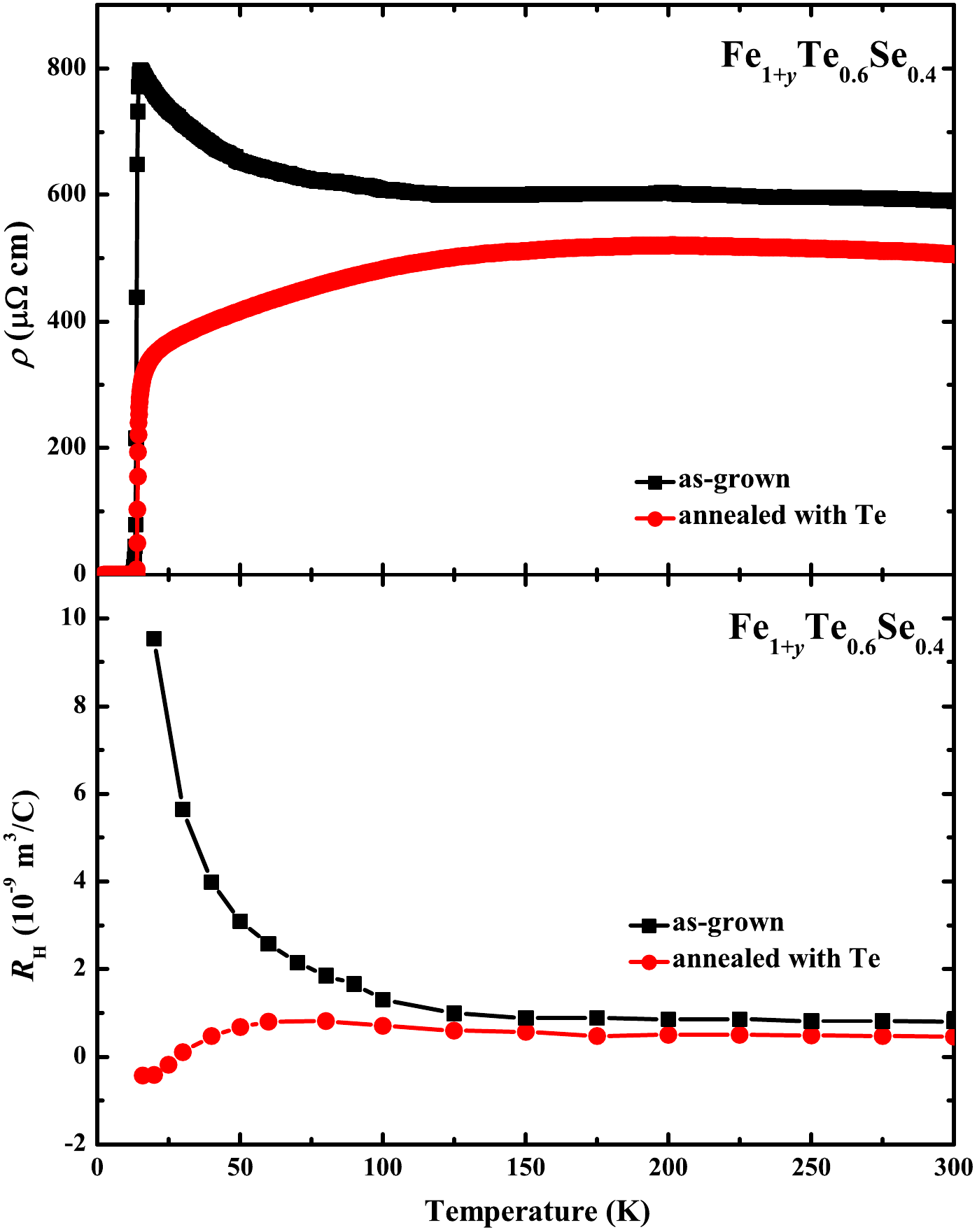}\\
　　\caption{(Color online) Temperature dependence of (a) in-plane resistivities (b) Hall coefficients for the as-grown and Te annealed Fe$_{1+y}$Te$_{0.6}$Se$_{0.4}$ single crystal. }\label{}
\end{figure}

Fig. 4(a) shows the temperature dependence of in-plane resistivity for the as-grown and annealed Fe$_{1+y}$Te$_{0.6}$Se$_{0.4}$ single crystals. For the as-grown sample, the resistivity maintains a nearly constant value at higher temperatures, followed by a semiconducting behavior (d$\rho$/d$T$ $<$ 0) below 120 K down to the superconducting transition. This semiconducting behavior is suppressed by Te annealing and replaced by a metallic behavior (d$\rho$/d$T$ $>$ 0). This is similar to that reported in O$_2$-annealed Fe$_{1+y}$Te$_{0.6}$Se$_{0.4}$ single crystal,\cite{23} and can be explained by the charge carrier delocalization coming from the removal of excess Fe by Te annealing. To further confirm the change in the behavior of charge carriers, Hall coefficient \emph{R}$_H$ of both the as-grown and annealed samples were measured and shown in Fig 4(b). Hall coefficients keep almost temperature independent values above 100 K in both the as-grown and annealed crystals. On the other hand, below about 100 K, an obvious divergence in \emph{R}$_H$ is observed. For the as-grown crystal, \emph{R}$_H$ increases with decreasing temperature. On the other hand, \emph{R}$_H$ in the annealed crystal keeps nearly temperature independent value above 50 K, followed by a sudden decrease, and finally changes sign from positive to negative before approaching \emph{T}$_c$. Strong temperature dependence of \emph{R}$_H$ with even a sign reversal is usually attributed to the multiband effect.\cite{24} The upturn in temperature dependent \emph{R}$_H$ in the as-grown sample may come from the effect of excess Fe, which is magnetic and interacts with the plane Fe magnetism proved by density functional study.\cite{25} The magnetic moment provided by excess Fe may localize the charge carriers, and cause the upturn in \emph{R}$_H$.\cite{26}

\begin{figure}\center

　　\includegraphics[width=8cm]{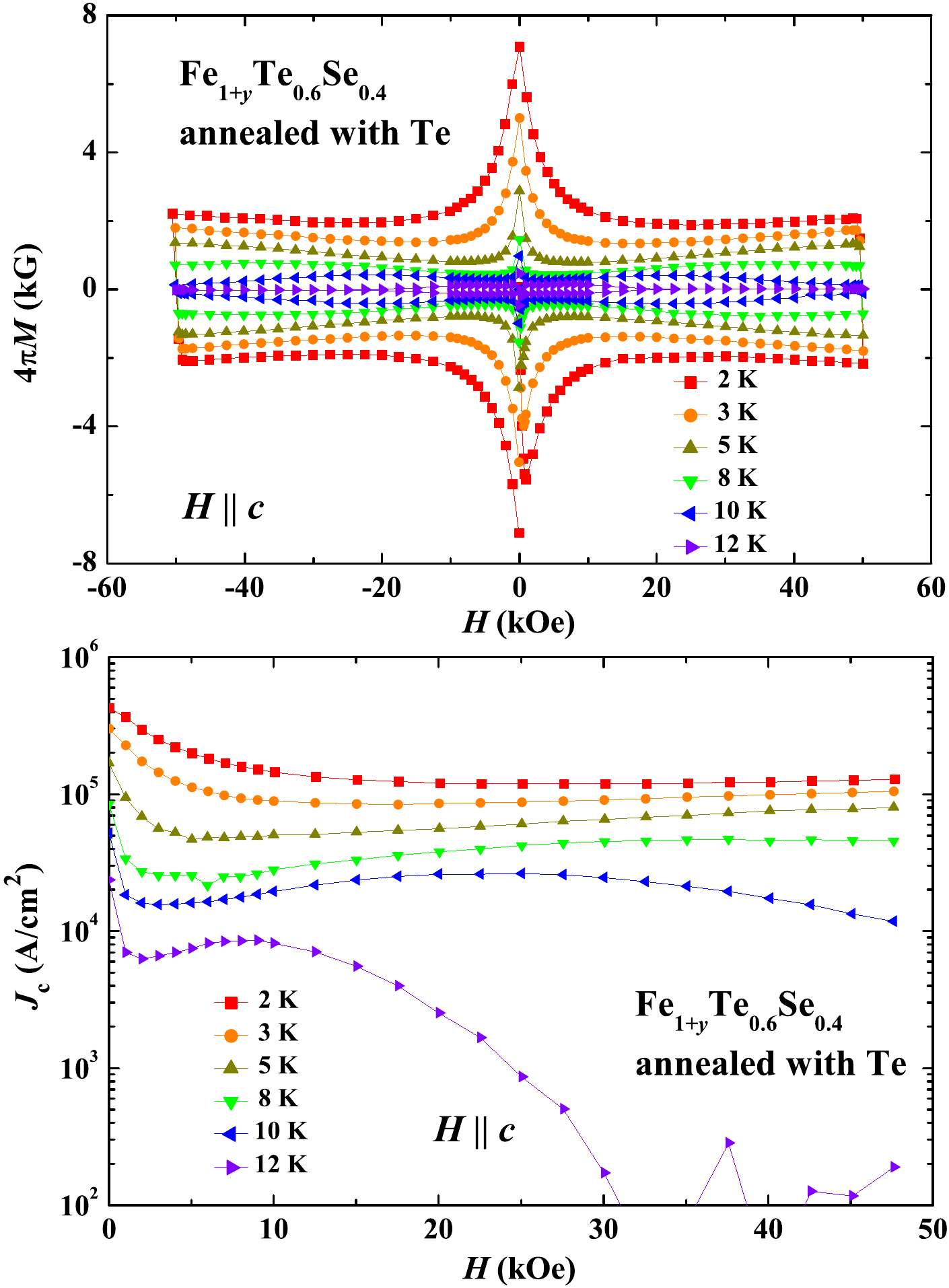}\\
　　\caption{(Color online) Magnetic hysteresis loops of Te annealed Fe$_{1+y}$Te$_{0.6}$Se$_{0.4}$ at different temperatures ranging from 2 to 12 K for \emph{H} $\|$ \emph{c}. (b) Magnetic field dependence of critical current densities for \emph{H} $\|$ \emph{c}.}\label{}
\end{figure}

Fig. 5(a) shows the MHLs of the annealed sample at temperature ranging from 2 to 12 K for \emph{H} $\|$ \emph{c}. A second magnetization peak (SMP), also known as the fish-tail effect (FE), can be witnessed, although not very prominent because the peak position is beyond the maximum applied field below 5 K. This can be seen more clearly in magnetic field dependent \emph{J}$_c$, which is calculated using eq. (1), and shown in Fig 5(b). Self-field \emph{J}$_c$ at 2 K is estimated as $\sim$ 4.3 $\times$ 10$^5$ A/cm$^2$, which is the largest among those reported in Fe(Te,Se) bulk samples.\cite{5, 17, 23, 27, 28, 29} Besides, the \emph{J}$_c$ is robust under applied field.
\begin{figure}\center

　　\includegraphics[width=8cm]{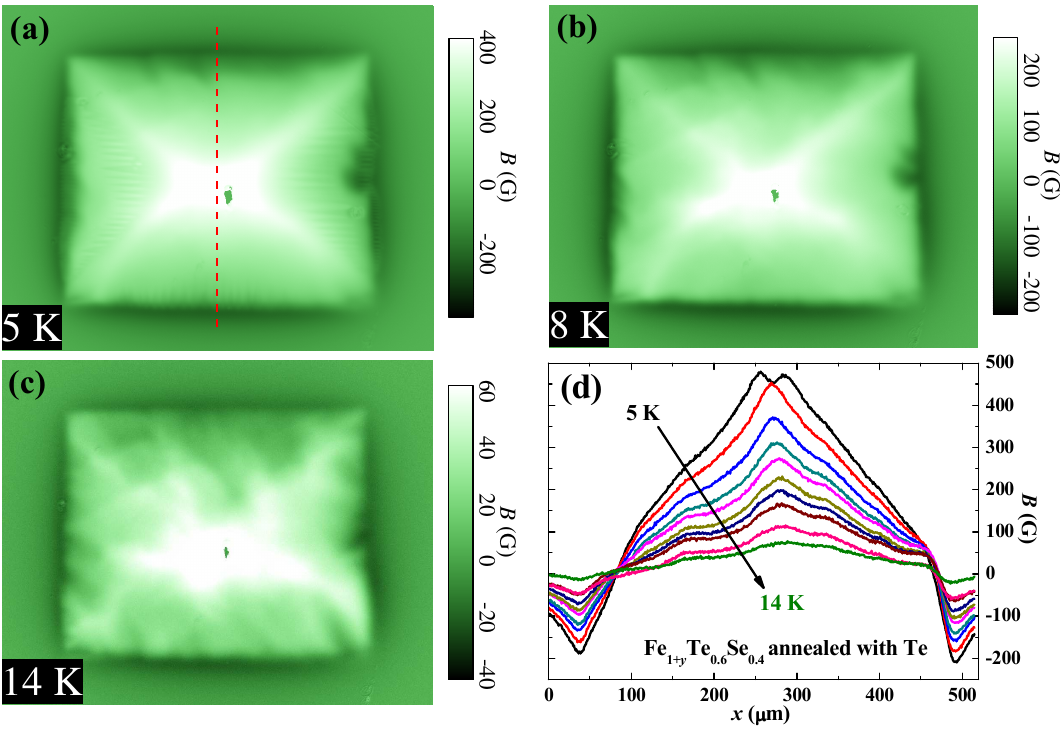}\\

　　\caption{(Color online) Magneto-optical (MO) images of the remnant states for an Te annealed Fe$_{1+y}$Te$_{0.6}$Se$_{0.4}$ single crystal at (a) 5 K, (b) 8 K, (c) 14 K. (d) The local magnetic induction profiles at temperatures from 5 K to 14 K taken along the dashed lines in (a). }\label{}
\end{figure}
The above estimation of \emph{J}$_c$ using the Bean model relies on the assumption that homogeneous current flows within the sample. To examine this assumption, we took MO images of the annealed sample in the remanent state at temperatures ranging from 5 to 14 K. This state is prepared by applying 800 Oe along the \emph{c}-axis for 1 s and removing it after zero-field cooling. Typical MO images at 5, 8, and 14 K are shown in Figs. 6(a) - (c), respectively. The MO image at 5 K shows clear roof-top patterns, indicating homogeneous current flowing in the crystal. The roof-top pattern becomes weaker at higher temperatures. Fig. 6(d) shows profiles of the magnetic induction along the dashed line at different temperatures shown in Fig. 6(a). From this profile, \emph{J}$_c$ can be roughly estimated by \emph{J}$_c$ $\sim$ $\Delta$\emph{B}/\emph{d}, where $\Delta$\emph{B} is the trapped field in the crystal, and \emph{d} is the thickness of the sample. \emph{J}$_c$ at 5 K is estimated as $\sim$ 3 $\times$ 10$^5$ A/cm$^2$, which is similar to that obtained from MHLs. The large and homogeneous critical current density obtained by MHLs and MO proves that Te annealing is effective to induce bulk superconductivity in Fe$_{1+y}$Te$_{0.6}$Se$_{0.4}$, which is promising for future applications in preparing wires and tapes.

In summary, we have systemically studied the effect of Te annealing in Fe$_{1+y}$Te$_{0.6}$Se$_{0.4}$ single crystal. \emph{T}$_c$ and \emph{J}$_c$ of the crystal are gradually enhanced by annealing with increasing amount of Te, until the molar ratio of Te to the sample reaches $\sim$ 0.1. Beyond this value, they are gradually reduced by annealing with more Te. Resistivity and Hall coefficient measurements show that annealing in Te vapor delocalizes the charge carriers by removing the excess Fe. EDX and XRD measurements show that the surface layers of the annealed crystal is composed of FeTe$_2$, which may come from the reaction of excess Fe with Te vapor. Self-field \emph{J}$_c$ of the annealed sample reaches a very high value $\sim$ 4.3 $\times$ 10$^5$ A/cm$^2$, and MO images reveal that the critical current density is homogeneously distributed in the crystal.

\begin{acknowledgment}

\verb|Acknowledgments|
This work was partly supported by the Natural Science Foundation of China, the Ministry of Science and Technology of China (973 project: No. 2011CBA00105), and the Japan-China Bilateral Joint Research Project by the Japan Society for the Promotion of Science.
\end{acknowledgment}

$^{*}$zxshi@seu.edu.cn   $^{\dag}$tamegai@ap.t.u-tokyo.ac.jp


\begin{thebibliography}{9}
\bibitem{1} Y. Kamihara, T. Watanabe, M. Hirano, and H. Hosono: J. Am. Chem. Soc. \textbf{130} (2008) 3296.
\bibitem{2} G. R. Stewart: Rev. Mod. Phys. \textbf{83} (2011) 1589.
\bibitem{3} F. C. Hsu, J. Y. Luo, K. W. Yeh, T. K. Chen, T. W. Huang, P. M. Wu, Y. C. Lee, Y. L. Huang, Y. Y. Chu, D. C. Yan, and M. K. Wu: Proc. Nat. Acad. Sci. \textbf{105} (2008) 14262.
\bibitem{4} B. C. Sales, A. S. Sefat, M. A. McGuire, R. Y. Jin, D. Mandrus, and Y. Mozharivskyj: Phys. Rev. B \textbf{79} (2009) 094521.
\bibitem{5} T. Taen, Y. Tsuchiya, Y. Nakajima, and T. Tamegai: Phys. Rev. B \textbf{80} (2009) 092502.
\bibitem{6} S. Margadonna, Y. Takabayashi, Y. Ohishi, Y. Mizuguchi, Y. Takano, T. Kagayama, T. Nakagawa, M. Takata, and K. Prassides: Phys. Rev. B \textbf{80} (2009) 064506.
\bibitem{7} S. Medvedev, T. M. McQueen, I. A. Troyan, T. Palasyuk, M. I. Eremets, R. J. Cava, S. Naghavi, F. Casper, V. Ksenofontov, G. Wortmann, and C. Felser: Nature Mater. \textbf{8} (2009) 630.
\bibitem{8} J. Guo, S. Jin, G. Wang, S. Wang, K. Zhu, T. Zhou, M. He, and X. Chen: Phys. Rev. B \textbf{82} (2010) 180520.
\bibitem{9} M. Burrard-Lucas, D. G. Free, S. J. Sedlmaier, J. D. Wright, S. J. Cassidy, Y. Hara, A. J. Corkett, T. Lancaster, P. J. Baker, S. J. Blundell, and S. J. Clarke: Nat Mater \textbf{12} (2013) 15.
\bibitem{10}  L. Sun, X. J. Chen, J. Guo, P. Gao, Q. Z. Huang, H. Wang, M. Fang, X. Chen, G. Chen, Q. Wu, C. Zhang, D. Gu, X. Dong, L. Wang, K. Yang, A. Li, X. Dai, H. K. Mao, and Z. Zhao: Nature \textbf{483} (2012) 67.
\bibitem{11} S. He, J. He, W. Zhang, L. Zhao, D. Liu, X. Liu, D. Mou, Y. B. Ou, Q. Y. Wang, Z. Li, L. Wang, Y. Peng, Y. Liu, C. Chen, L. Yu, G. Liu, X. Dong, J. Zhang, C. Chen, Z. Xu, X. Chen, X. Ma, Q. Xue, and X. J. Zhou: Nat Mater \textbf{12} (2013) 605.
\bibitem{12} W. Bao, Y. Qiu, Q. Huang, M. A. Green, P. Zajdel, M. R. Fitzsimmons, M. Zhernenkov, S. Chang, M. Fang, B. Qian, E. K. Vehstedt, J. Yang, H. M. Pham, L. Spinu, and Z. Q. Mao: Phys. Rev. Lett. \textbf{102} (2009) 247001.
\bibitem{13} M. Bendele, P. Babkevich, S. Katrych, S. N. Gvasaliya, E. Pomjakushina, K. Conder, B. Roessli, A. T. Boothroyd, R. Khasanov, and H. Keller: Phys. Rev. B \textbf{82} (2010) 212504.
\bibitem{14} T. M. McQueen, Q. Huang, V. Ksenofontov, C. Felser, Q. Xu, H. Zandbergen, Y. S. Hor, J. Allred, A. J. Williams, D. Qu, J. Checkelsky, N. P. Ong, and R. J. Cava: Phys. Rev. B \textbf{79} (2009) 014522.
\bibitem{15} D. J. Singh: Sci. Technol. Adv. Mat. \textbf{13} (2012) 054304.
\bibitem{16} K. Tanabe and H. Hosono: Jpn. J. Appl. Phys. \textbf{51} (2012) 010005.
\bibitem{17} Y. Sun, T. Taen, Y. Tsuchiya, Z. X. Shi, and T. Tamegai: Supercond. Sci. Technol. \textbf{26} (2013) 015015.
\bibitem{18} Y. Koshika, T. Usui, S. Adachi, T. Watanabe, K. Sakano, S. Simayi, and M. Yoshizawa: J. Phys. Soc. Jpn. \textbf{82} (2013) 023703.
\bibitem{19} A. Soibel, E. Zeldov, M. Rappaport, Y. Myasoedov, T. Tamegai, S. Ooi, M. Konczykowski, and V. B. Geshkenbein: Nature \textbf{406} (2000) 282.
\bibitem{20} C. P. Bean: Rev. Mod. Phys. \textbf{36} (1964) 31.
\bibitem{21} F. Gronvold, H. Haraldsen, and J. Vihovde: Acta. Chem. Scand. \textbf{8} (1954) 1927.
\bibitem{22} J. B. Ward, V. H. McCann, J. H. T. Bates, and M. P. L. Quin: J. of Phys. C: Solid State Phys. \textbf{11} (1978) 377.
\bibitem{23} Y. Sun, T. Taen, Y. Tsuchiya, Q. Ding, S. Pyon, Z. Shi, and T. Tamegai: Appl. Phys. Express \textbf{6} (2013) 043101.
\bibitem{24} F. Chen, B. Zhou, Y. Zhang, J. Wei, H. W. Ou, J. F. Zhao, C. He, Q. Q. Ge, M. Arita, K. Shimada, H. Namatame, M. Taniguchi, Z. Y. Lu, J. Hu, X. Y. Cui, and D. L. Feng: Phys. Rev. B \textbf{81} (2010) 014526.
\bibitem{25} L. Zhang, D. J. Singh, and M. H. Du: Phys. Rev. B \textbf{79} (2009) 012506.
\bibitem{26} T. J. Liu, X. Ke, B. Qian, J. Hu, D. Fobes, E. K. Vehstedt, H. Pham, J. H. Yang, M. H. Fang, L. Spinu, P. Schiffer, Y. Liu, and Z. Q. Mao: Phys. Rev. B \textbf{80} (2009) 174509.
\bibitem{27} C. S. Yadav and P. L. Paulose: New. J. Phys. \textbf{11} (2009) 103046.
\bibitem{28} Y. Liu, R. K. Kremer, and C. T. Lin: Europhys. Lett. \textbf{92} (2010) 57004.
\bibitem{29} T. Taen, Y. Tsuchiya, Y. Nakajima, and T. Tamegai: Physica C \textbf{470} (2010) 1106.


\end{thebibliography}
\end{document}